\documentclass{PoS}
\usepackage{axodraw,amsmath,alltt,feynarts,pifont}

\DeclareSymbolFont{ttoperators}{OT1}{cmtt}{m}{n}
\newcommand\Code[1]{{%
  \mathcode`\"="0\the\symttoperators22%
  \mathchardef\$="4\the\symttoperators24%
  \mathcode`\(="4\the\symttoperators28%
  \mathcode`\)="5\the\symttoperators29%
  \mathcode`\/="0\the\symttoperators2F%
  \mathcode`\[="4\the\symttoperators5B%
  \mathcode`\]="5\the\symttoperators5D%
  \mathchardef\{="4\the\symttoperators7B%
  \mathchardef\}="5\the\symttoperators7D%
  \ensuremath{\mathtt{#1}}}}
\newcommand\Var[1]{{\ensuremath{\mathit{#1}}}}

\newcommand\lbrac{\symbol{123}}
\newcommand\rbrac{\symbol{125}}
\newcommand\Brac[1]{\lbrac#1\rbrac}
\newcommand\unity{\mathrm{\rm 1\mskip-4.5mu l}}

\newcommand\eg{e.g.\ }
\newcommand\ie{i.e.\ }

\newcommand\sq{$^{\text{2}}$}
\newcommand\ket[1]{\left| #1\right\rangle}
\newcommand\bra[1]{\left\langle #1\right|}
\newcommand\Tr{\mathop{\mathrm{Tr}}}
\renewcommand\Re{\mathop{\mathrm{Re}}}

\newcommand\M{\mathcal{M}}
\newcommand\Mtree{\ensuremath{\mathcal{M}_{\mathrm{tree}}}}
\newcommand\Mloop{\ensuremath{\mathcal{M}_{\mathrm{loop}}}}

\title{Feynman Diagram Calculations with \\
FeynArts, FormCalc, and LoopTools}

\ShortTitle{FeynArts, FormCalc, and LoopTools}

\author{\speaker{Thomas Hahn} \\
        Max-Planck-Institut f\"ur Physik \\
	F\"ohringer Ring 6 \\
	D--80805 Munich, Germany \\
        E-mail: \email{hahn@feynarts.de}}

\abstract{This article describes the latest versions of the Mathematica 
packages FeynArts, FormCalc, and LoopTools for the generation and 
evaluation of one-loop diagrams.}

\FullConference{13th International Workshop on Advanced Computing and 
Analysis Techniques in Physics Research\\
                 February 22-27, 2010\\
                 Jaipur, India}

\begin{document}

\section{Introduction}

FeynArts \cite{FeynArts}, FormCalc, and LoopTools \cite{FormCalc} are 
programs for the generation and calculation of Feynman diagrams.  The 
`canonical' problem they solve is the computation of the cross-section 
up to one-loop order for a given scattering reaction, in a highly 
automated way.  Mathematica as the basic programming language makes it 
straightforward to use in particular intermediate results in a variety of 
ways, for example checks of Ward identities or the extraction of Wilson 
coefficients.  This note gives an overview of the system with emphasis 
on recently added features.


\section{FeynArts}

FeynArts is a Mathematica package for the generation and visualization 
of Feynman diagrams and amplitudes.  Launched in 1991 \cite{FeynArts1}, 
the current version 3.5 still uses almost the same syntax, though with 
many extensions.  The generation of amplitudes is a three-step process.  
In the first step, the distinct topologies for a given number of loops 
and external legs are produced, \eg
\begin{alltt}
   top = CreateTopologies[1, 1 \(\to\) 2]
\end{alltt}
This is a purely geometrical/topological task and requires no physics 
input.  The internal algorithm starts from given zero-leg topologies of 
the requested loop order and successively adds legs.

In the second step, the model's particle content is read from a Model 
File and the fields are distributed over the topologies in all 
admissible ways, \eg
\begin{alltt}
   ins = InsertFields[top, F[4,\Brac{3}] \(\to\) \Brac{F[4,\Brac{2}], V[1]}]
\end{alltt}
Finally, the Feynman rules are applied with
\begin{verbatim}
   amp = CreateFeynAmp[ins]
\end{verbatim}
The field labelling above is the one of the default model, 
\Code{SM.mod}, and corresponds to the decay $b\to s\gamma$, where $b$ 
and $s$ are the third and second members of the down-type quark class 
\Code{F[4]}, and \Code{V[1]} is the photon.  This notation is part of 
the more general concept of field levels:
\begin{itemize}
\item
The Generic Level determines the space--time properties of a field, \eg 
a fermion \Code{F}.  It also fixes the kinematic properties of the 
couplings.  For example, the \Code{FFS} coupling is of the form 
$G_+\omega_+ + G_-\omega_-$, where $\omega_\pm = (\unity\pm\gamma_5)/2$, 
with coefficients $G_\pm$ that depend on model parameters only.

\item
The Classes Level specifies the particle up to `simple' index 
substitutions, \eg the down-type quark class \Code{F[4]} (where the 
generation index is not yet given).

\item
The Particles Level spells out any indices left unspecified, \eg 
the bottom quark \Code{F[4,\{3\}]}.
\end{itemize}
The reason for this splitting is mainly economy: kinematic 
simplifications can be performed at Generic Level, where there are 
typically much fewer diagrams than at lower levels.  Likewise, `trivial' 
sums \eg over fermion generations need not be written out explicitly in 
terms of Particles-Level Feynman diagrams.

The diagrams returned by \Code{CreateTopologies} and \Code{InsertFields} 
can be drawn with \Code{Paint}, with output as Mathematica Graphics 
object, PostScript, or \LaTeX.  \LaTeX\ code produced by \Code{Paint} 
can be post-processed (\eg `touched up' for publication) with the 
FeynEdit editor \cite{FeynEdit}.

A diagram in the output is encoded as
\Code{FeynAmp[\Var{id},\,\Var{loopmom},\,\Var{genamp},\,\Var{ins}]}.
For illustration, consider the diagram
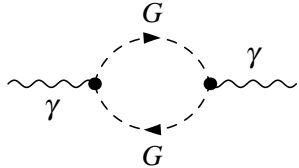
\begin{center}
\vspace*{-7ex}
\begin{feynartspicture}(120,120)(1,1)
\FADiagram{}
\FAProp(0.,10.)(6.,10.)(0.,){Sine}{0}
\FALabel(3.,8.93)[t]{$\gamma$}
\FAProp(20.,10.)(14.,10.)(0.,){Sine}{0}
\FALabel(17.,11.07)[b]{$\gamma$}
\FAProp(6.,10.)(14.,10.)(0.8,){ScalarDash}{-1}
\FALabel(10.,5.73)[t]{$G$}
\FAProp(6.,10.)(14.,10.)(-0.8,){ScalarDash}{1}
\FALabel(10.,14.27)[b]{$G$}
\FAVert(6.,10.){0}
\FAVert(14.,10.){0}
\end{feynartspicture}
\vspace*{-6ex}
\end{center}
\begin{itemize}
\item
\Var{id} is an identifier for bookkeeping, \eg 
\Code{GraphID[Topology == 1, Generic == 1]},

\item
\Var{loopmom} identifies the loop momenta in the form
\Code{Integral[q1]},

\item
\Var{genamp} is the generic amplitude,

\Code{\dfrac{I}{32~Pi^4}~RelativeCF}~\ding{192} \\[1ex]
\Code{FeynAmpDenominator[\dfrac{1}{q1^2 - Mass[S[Gen3]]^2},
  \dfrac{1}{(-p1 + q1)^2 - Mass[S[Gen4]]^2}]}~\ding{193} \\[1ex]
\Code{(p1 - 2\,q1)[Lor1]~~(-p1 + 2\,q1)[Lor2]}~\ding{194} \\[1ex]
\Code{ep[V[1],p1,Lor1]~~ep^*[V[1],k1,Lor2]}~\ding{195} \\[1ex]
\Code{G^{(0)}_{SSV}[(Mom[1]-Mom[2])[KI1[3]]]~~
      G^{(0)}_{SSV}[(Mom[1]-Mom[2])[KI1[3]]]}~\ding{196}

where individual items can easily be identified: prefactor \ding{192}, 
loop denominators \ding{193}, coupling structure \ding{194}, 
polarization vectors \ding{195}, coupling constants \ding{196}.

\item
\Var{ins} is a list of rules substituting the unspecified items in the 
generic amplitude,

\Code{\{~Mass[S[Gen3]],~Mass[S[Gen4]],} \\
\Code{\hphantom{\{~}G^{(0)}_{SSV}[(Mom[1]-Mom[2])[KI1[3]]],} \\
\Code{\hphantom{\{~}G^{(0)}_{SSV}[(Mom[1]-Mom[2])[KI1[3]]],~RelativeCF~\}~\to} \\
\Code{Insertions[Classes][\{MW,~MW,~I\ EL,~-I\ EL,~2\}] }
\end{itemize}


\subsection{Model Files}

The Model Files are ordinary Mathematica text files loaded by FeynArts 
during model initialization.  They supply certain objects, \eg
\Code{M\$ClassesDescription}, the list of particles, and 
\Code{M\$CouplingMatrices}, the list of couplings.  Generic 
(\Code{.gen}) and Classes (\Code{.mod}) Model Files store the kinematic 
and constant part of the coupling, respectively.

FeynArts further distinguishes Basic and Partial (Add-On) Model Files. 
Basic Model Files, such as \Code{SM.mod}, \Code{MSSM.mod}, can be 
modified by Add-On Model Files, as in
\begin{alltt}
   InsertFields[..., Model \(\to\) \Brac{"MSSMQCD", "FV"}]
\end{alltt}
This loads the Basic Model File \Code{MSSMQCD.mod} and modifies it 
through the Add-On Model File \Code{FV.mod} (non-minimal flavour 
violation).  The brace notation works similarly for Generic Model 
files.  The Add-On Model File typically modifies (rather than
overwrites) its objects.

Model Files for FeynArts can currently be generated by FeynRules
\cite{FeynRules} and LanHEP \cite{LanHEP}.  The SARAH package
\cite{SARAH} is useful for the high-level derivation of SUSY models.  
FeynArts itself includes the ModelMaker tool which turns a suitably 
defined Lagrangian into a Model File.  For further details of model 
construction the reader is referred to the respective manuals.


\subsection{Linear Combination of Fields}

Starting from Version 3.5, FeynArts can automatically linear-combine 
fields, \ie one can specify the couplings in terms of gauge rather than 
mass eigenstates.  For example:
\begin{small}
\begin{alltt}
   M$ClassesDescription = \lbrac ...,
     F[11] == \lbrac...,
       Indices \(\to\) \Brac{Index[Neutralino]},
       Mixture \(\to\) ZNeu[Index[Neutralino],1] F[111] +
                  ZNeu[Index[Neutralino],2] F[112] +
                  ZNeu[Index[Neutralino],3] F[113] +
                  ZNeu[Index[Neutralino],4] F[114]\rbrac \rbrac
\end{alltt}
\end{small}
\Code{M\$CouplingMatrices} may now be given in terms of the gauge 
eigenstates \Code{F[111]}\dots\Code{F[114]}, which makes the expressions 
much shorter.  \Code{F[111]}\dots\Code{F[114]} drop out completely after
model initialization, however, as they are not themselves listed in 
\Code{M\$ClassesDescription}.

Higher-order mixings can be added, too:
\begin{small}
\begin{alltt}
   M$ClassesDescription = \lbrac ...,
     S[1] == \Brac{...},
     S[2] == \Brac{...},
     S[10] == \lbrac...,
       Indices \(\to\) \Brac{Index[Higgs]},
       Mixture \(\to\) UHiggs[Index[Higgs],1] S[1] +
                  UHiggs[Index[Higgs],2] S[2],
       InsertOnly \(\to\) \Brac{External, Internal}\rbrac \rbrac
\end{alltt} 
\end{small}
This time, \Code{S[10]} \emph{and} \Code{S[1]},\,\Code{S[2]} appear in 
\Code{M\$ClassesDescription} and hence the coupling list contains both 
mixed and unmixed states, in all possible combinations.  Due to the 
\Code{InsertOnly}, \Code{S[10]} is inserted only on tree-level parts of 
the diagram, not in loops.


\subsection{Enhanced Diagram Selection}

In recent FeynArts versions, many functions have been added or extended 
to ease diagram selection: \Code{DiagramSelect}, \Code{DiagramGrouping}, 
\Code{DiagramMap}, \Code{DiagramComplement}.
Also new or extended are many `filter functions' which simplify the 
construction of sophisticated filters for the selection functions above: 
\Code{Vertices}, \Code{FieldPoints}, \Code{FermionRouting}, 
\Code{FeynAmpCases}, \Code{FieldMatchQ}, \Code{FieldMemberQ}, 
\Code{FieldPointMatchQ}, \Code{FieldPointMemberQ}.

\def\diagding#1{\raise 2ex\hbox{#1}}
\def\diagyes{\diagding{\Green{\ding{52}}}}
\def\diagno{\diagding{\Red{\ding{56}}}}

To pick just two examples: the selection of wave-function corrections 
(WFc) has become more flexible.  The exclusion of WFc can be specified 
individually for every external leg,
\begin{alltt}
   CreateTopologies[..., ExcludeTopologies \(\to\) WFCorrections[1|3]]
\end{alltt}
The filter function \Code{WFCorrectionFields} returns the in- and 
out-fields of the self-energy constituting the WFc.  It solves the 
problem of treating WFc with same outer particles (usually omitted)
and different particles (kept unless some on-shell scheme is employed) 
differently, \eg
\begin{verbatim}
   DiagramSelect[..., UnsameQ@@ WFCorrectionFields[##] &]
\end{verbatim}

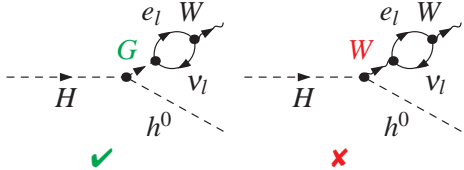
\vspace*{-2ex}

\begin{feynartspicture}(180,90)(2,1)
\FADiagram{\kern -1em\diagyes}
\FAProp(0.,10.)(11.,10.)(0.,){ScalarDash}{1}
\FALabel(5.5,8.93)[t]{$H$}
\FAProp(20.,15.)(17.3,13.5)(0.,){Sine}{-1}
\FALabel(18.3773,15.1249)[br]{$W$}
\FAProp(20.,5.)(11.,10.)(0.,){ScalarDash}{0}
\FALabel(15.3487,6.8436)[tr]{$h^0$}
\FAProp(11.,10.)(13.7,11.5)(0.,){ScalarDash}{1}
\FALabel(12.0773,11.6249)[br]{\Green{$G$}}
\FAProp(17.3,13.5)(13.7,11.5)(-0.8,){Straight}{1}
\FALabel(16.5727,10.1851)[tl]{$\nu_l$}
\FAProp(17.3,13.5)(13.7,11.5)(0.8,){Straight}{-1}
\FALabel(14.4273,14.8149)[br]{$e_l$}
\FAVert(11.,10.){0}
\FAVert(17.3,13.5){0}
\FAVert(13.7,11.5){0}

\FADiagram{\kern -1em\diagno}
\FAProp(0.,10.)(11.,10.)(0.,){ScalarDash}{1}
\FALabel(5.5,8.93)[t]{$H$}
\FAProp(20.,15.)(17.3,13.5)(0.,){Sine}{-1}
\FALabel(18.3773,15.1249)[br]{$W$}
\FAProp(20.,5.)(11.,10.)(0.,){ScalarDash}{0}
\FALabel(15.3487,6.8436)[tr]{$h^0$}
\FAProp(11.,10.)(13.7,11.5)(0.,){Sine}{1}
\FALabel(12.0773,11.6249)[br]{\Red{$W$}}
\FAProp(17.3,13.5)(13.7,11.5)(-0.8,){Straight}{1}
\FALabel(16.5727,10.1851)[tl]{$\nu_l$}
\FAProp(17.3,13.5)(13.7,11.5)(0.8,){Straight}{-1}
\FALabel(14.4273,14.8149)[br]{$e_l$}
\FAVert(11.,10.){0}
\FAVert(17.3,13.5){0}
\FAVert(13.7,11.5){0}
\end{feynartspicture}

The new filter function \Code{FermionRouting} can be used to
select diagrams according to their fermion structure,  \eg
\begin{verbatim}
   DiagramSelect[..., FermionRouting[##] === {1,3, 2,4} &]
\end{verbatim}
selects only diagrams where external legs 1--3 and 2--4 are connected
through fermion lines.

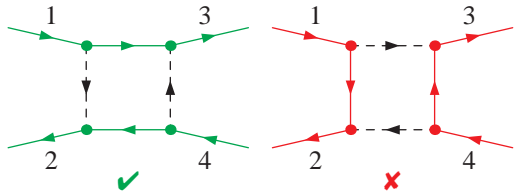
\vspace*{-2ex}

\begin{feynartspicture}(200,100)(2,1)
\FADiagram{\diagyes}
\FAProp(6.5,13.5)(6.5,6.5)(0.,){ScalarDash}{1}
\FAProp(13.5,13.5)(13.5,6.5)(0.,){ScalarDash}{-1}
\Green{%
\FAProp(0.,15.)(6.5,13.5)(0.,){Straight}{1}
\FAProp(0.,5.)(6.5,6.5)(0.,){Straight}{-1}
\FAProp(20.,15.)(13.5,13.5)(0.,){Straight}{-1}
\FAProp(20.,5.)(13.5,6.5)(0.,){Straight}{1}
\FAProp(6.5,13.5)(13.5,13.5)(0.,){Straight}{1}
\FAProp(6.5,6.5)(13.5,6.5)(0.,){Straight}{-1}
\FAVert(6.5,13.5){0}
\FAVert(6.5,6.5){0}
\FAVert(13.5,13.5){0}
\FAVert(13.5,6.5){0}
}
\FALabel(3.59853,15.2803)[b]{\small $1$}
\FALabel(3.59853,4.71969)[t]{\small $2$}
\FALabel(16.4015,15.2803)[b]{\small $3$}
\FALabel(16.4015,4.71969)[t]{\small $4$}

\FADiagram{\diagno}
\FAProp(6.5,13.5)(13.5,13.5)(0.,){ScalarDash}{1}
\FAProp(6.5,6.5)(13.5,6.5)(0.,){ScalarDash}{-1}
\Red{%
\FAProp(0.,15.)(6.5,13.5)(0.,){Straight}{1}
\FAProp(0.,5.)(6.5,6.5)(0.,){Straight}{-1}
\FAProp(20.,15.)(13.5,13.5)(0.,){Straight}{-1}
\FAProp(20.,5.)(13.5,6.5)(0.,){Straight}{1}
\FAProp(6.5,13.5)(6.5,6.5)(0.,){Straight}{1}
\FAProp(13.5,13.5)(13.5,6.5)(0.,){Straight}{-1}
\FAVert(6.5,13.5){0}
\FAVert(6.5,6.5){0}
\FAVert(13.5,13.5){0}
\FAVert(13.5,6.5){0}
}
\FALabel(3.59853,15.2803)[b]{\small $1$}
\FALabel(3.59853,4.71969)[t]{\small $2$}
\FALabel(16.4015,15.2803)[b]{\small $3$}
\FALabel(16.4015,4.71969)[t]{\small $4$}
\end{feynartspicture}


\section{FormCalc}

The output of FeynArts is not directly in a good shape for numerical 
evaluation.  It contains uncontracted indices, unregularized loop 
integrals, fermion traces, SU($N$) generators, etc.  The symbolic 
expressions for the diagrams are thus first simplified algebraically 
with FormCalc, which performs the following steps: indices are 
contracted, fermion traces evaluated, open fermion chains simplified, 
colour structures standardized, tensor integrals reduced, abbreviations 
introduced.

Most of these steps are internally executed in FORM \cite{FORM}, a 
computer-algebra system whose instruction set has many adaptations 
especially useful in high-energy physics.  The interfacing with FORM is 
transparent to the user, \ie the user does not have to work with the 
FORM code.  FormCalc thus combines the speed of FORM with the powerful 
instruction set of Mathematica and the latter greatly facilitates 
further processing of the results.

The main function is \Code{CalcFeynAmp} which is applied to a FeynArts 
amplitude (the output of \Code{CreateFeynAmp}) and combines the steps 
outlined above.  Its output is in general a linear combination of loop 
integrals with prefactors that contain model parameters, kinematic 
variables, and abbreviations introduced by FormCalc, \eg
\begin{alltt}
   C0i[cc0, MW2, MW2, S, MW2, MZ2, MW2] *
     ( -4 Alfa2 CW2 MW2/SW2 S AbbSum16 +
       32 Alfa2 CW2/SW2 S\sq AbbSum28 +
       4 Alfa2 CW2/SW2 S\sq AbbSum30 -
       8 Alfa2 CW2/SW2 S\sq AbbSum7 +
       Alfa2 CW2/SW2 S\,(T\,-\,U) Abb1 +
       8 Alfa2 CW2/SW2 S\,(T\,-\,U) AbbSum29 )
\end{alltt}
The first line represents the one-loop integral $C_0(M_W^2, M_W^2, s, 
M_W^2, M_Z^2, M_W^2)$, multiplied with a linear combination of 
abbreviations like \Code{Abb1} or \Code{AbbSum29} with coefficients 
containing kinematical invariants like the Mandelstam variables 
\Code{S}, \Code{T}, and \Code{U} and model parameters such as 
$\Code{Alfa2} = \alpha^2$.


\subsection{Abbreviations}
\label{sect:abbr}

The automated introduction of abbreviations is a key concept in 
FormCalc.  It is crucial in rendering an amplitude as compact as 
possible.  The main effect comes from three layers of recursively 
defined abbreviations, introduced when the amplitude is read back from 
FORM, \ie during \Code{CalcFeynAmp}.  For example:
\begin{center}
\begin{picture}(200,60)(0,12)
\SetScale{.8}

\Text(0,60)[bl]{\Code{AbbSum29 = Abb2 + Abb22 + Abb23 + Abb3}}
\SetOffset(108,12)
\EBox(-19,55)(19,72)
\Line(-19,55)(-86,41)
\Line(-86,41)(-86,28)
\Line(19,55)(86,41)
\Line(86,41)(86,28)

\Text(0,24)[b]{\Code{Abb22 = Pair1~Pair3~Pair6}}
\SetOffset(129,12)
\EBox(-19,25)(19,42)
\Line(-19,25)(-82,13)
\Line(-82,13)(-82,0)
\Line(19,25)(82,13)
\Line(82,13)(82,0)

\Text(0,4)[]{\Code{Pair3 = Pair[e[3],\,k[1]]}}
\end{picture}
\end{center}
Written out, this abbreviation is equivalent to
\begin{small}
\begin{alltt}
   Pair[e[1],\,e[2]] Pair[e[3],\,k[1]] Pair[e[4],\,k[1]] +
   Pair[e[1],\,e[2]] Pair[e[3],\,k[2]] Pair[e[4],\,k[1]] +
   Pair[e[1],\,e[2]] Pair[e[3],\,k[1]] Pair[e[4],\,k[2]] +
   Pair[e[1],\,e[2]] Pair[e[3],\,k[2]] Pair[e[4],\,k[2]]
\end{alltt}
\end{small}
In addition to these abbreviations assigned by \Code{CalcFeynAmp},
FormCalc introduces another set of abbreviations for the loop integrals
when generating Fortran code.


\subsection{Categories}

Both of the aforementioned types of abbreviations, the latter in 
particular, are costly in CPU time.  It is thus key to performance that 
the abbreviations are grouped into different categories:
\begin{enumerate}
\item Abbreviations that depend on the helicities.
\item Abbreviations that depend on angular variables.
\item Abbreviations that depend only on $\sqrt s$.
\end{enumerate}
Correct execution of the different categories guarantees that almost no 
redundant evaluations are made, \eg in a $2\to 2$ process with external 
unpolarized fermions, statements in the innermost loop over the 
helicities are executed $2^4$ times as often as those in the loop over 
the angle.  This technique of moving invariant expressions out of the 
loop is known as `hoisting' in computer science.

The \Code{Abbreviate} function extends the advantages of the 
abbreviation system to arbitrary expressions.  Its usage is for example:
\begin{verbatim}
   abbrexpr = Abbreviate[expr, 5]
\end{verbatim}
The second argument, 5, specifies the level below which abbreviations 
are introduced, \ie how much of expression is `abbreviated away.'  In 
the extreme, for a level of 1, the result is just a single symbol. 
Abbreviationing also has the nice side effect that duplicate expressions 
are replaced by the same symbol.  This new type of abbreviations for 
subexpressions has to be retrieved separately from the other ones with 
\Code{Subexpr[]}.

Abbreviations and subexpressions from an earlier Mathematica session 
must be `registered' first using \Code{RegisterAbbr[\Var{abbr}]} and
\Code{RegisterSubexpr[\Var{subexpr}]}.


\subsection{Weyl chains and Dirac chains}

Amplitudes with external fermions have the form $\M = 
\sum_{i = 1}^n c_i\, F_i$, where the $F_i$ are (products of) fermion 
chains.  The textbook recipe is to compute probabilities, \eg
$|\M|^2 = \sum_{i, j = 1}^n c_i^*\, c_j\, F_i^* F_j$, and 
evaluate the $F_i^* F_j$ by trace techniques: 
$\left|\bra{u}\Gamma\ket{v}\right|^2 = 
\bra{u}\Gamma\ket{v}\bra{v}\bar\Gamma\ket{u} = 
\Tr\bigl(\Gamma\ket{v}\bra{v}\bar\Gamma\ket{u}\bra{u}\bigr)$.

The problem with this approach is that instead of $n$ of the $F_i$ one
needs to compute $n^2$ of the $F_i^* F_j$.  Since essentially $n\sim
(\text{number of vectors})!$, this quickly becomes a limiting factor in
problems involving many vectors, \eg in multi-particle final states or
polarization effects.

The solution is of course to compute the amplitude $\M$ directly and 
this is done most conveniently in the Weyl--van der Waerden formalism 
\cite{WvdW}.  The implementation of this technique in an automated 
program has been outlined in \cite{optimizations}.

The \Code{FermionChains} option of \Code{CalcFeynAmp} determines how 
fermion chains are returned: \Code{Weyl}, the default, selects Weyl 
chains.  \Code{Chiral} and \Code{VA} select Dirac chains in the chiral 
($\omega_+/\omega_-$) and vector/axial-vector ($\unity/\gamma_5$) 
decomposition, respectively.  The Weyl chains need not be further 
evaluated with \Code{HelicityME}, which applies the trace technique.

As numerical calculations are done mostly using Weyl chains therefore, 
there has been a paradigm shift for Dirac chains to make them better 
suited for analytical purposes, \eg the extraction of Wilson 
coefficients.

The \Code{FermionOrder} option of \Code{CalcFeynAmp} implements Fierz 
reordering, allowing the user to force Dirac chains into almost any 
desired order.  \Code{FermionOrder} does not only allow for explicit 
orderings, but can take the \Code{Colour} option, too, in which case the 
spinor indices are brought into the same order as the colour indices, a 
convention commonly found in the literature.

Antisymmetrized Dirac chains are chosen with the \Code{Antisymmetrize} 
option of \Code{CalcFeynAmp}.  They are indicated by a negative 
identifier, \eg $\Code{DiracChain[-1,\mu,\nu]} = \sigma_{\mu\nu}$.


\subsection{Alternate Link between FORM and Mathematica}

FORM is renowned for being able to handle very large expressions. To 
produce (pre)simplified expressions, however, terms have to be wrapped 
in functions, to avoid immediate expansion.  The number of terms 
\emph{in a function} is unfortunately rather limited in FORM: on 32-bit 
systems to 32767.  While FormCalc gets more sophisticated in 
pre-simplifying amplitudes, users want to compute larger amplitudes and 
have thus seen many `overflow' messages from FORM recently.

In FormCalc Versions 6 and up, the pre-simplified generic amplitude is 
intermediately sent to Mathematica for introducing abbreviations through 
FORM's external channels \cite{extform}.  This results in a significant 
reduction in size of intermediate expressions.

The following example is taken from the tree-level $uu\to gg$ amplitude.
The expression passed from FORM to Mathematica is
\begin{scriptsize}
\begin{verbatim}
    +Den[U,MU2]*(
      -8*SUNSum[Col5,3]*SUNT[Glu3,Col5,Col2]*SUNT[Glu4,Col1,Col5]*mul[Alfas*Pi]*
      abb[fme[WeylChain[DottedSpinor[k1,MU,-1],6,Spinor[k2,MU,1]]]*ec3.ec4
        -1/2*fme[WeylChain[DottedSpinor[k1,MU,-1],6,ec3,ec4,Spinor[k2,MU,1]]]
        +fme[WeylChain[DottedSpinor[k1,MU,-1],7,Spinor[k2,MU,1]]]*ec3.ec4
        -1/2*fme[WeylChain[DottedSpinor[k1,MU,-1],7,ec3,ec4,Spinor[k2,MU,1]]]]*MU
      -4*SUNSum[Col5,3]*SUNT[Glu3,Col5,Col2]*SUNT[Glu4,Col1,Col5]*mul[Alfas*Pi]*
      abb[fme[WeylChain[DottedSpinor[k1,MU,-1],6,ec3,ec4,k3,Spinor[k2,MU,1]]]
        -2*fme[WeylChain[DottedSpinor[k1,MU,-1],6,ec4,Spinor[k2,MU,1]]]*ec3.k2
        -2*fme[WeylChain[DottedSpinor[k1,MU,-1],6,k3,Spinor[k2,MU,1]]]*ec3.ec4
        +fme[WeylChain[DottedSpinor[k1,MU,-1],7,ec3,ec4,k3,Spinor[k2,MU,1]]]
        -2*fme[WeylChain[DottedSpinor[k1,MU,-1],7,ec4,Spinor[k2,MU,1]]]*ec3.k2
        -2*fme[WeylChain[DottedSpinor[k1,MU,-1],7,k3,Spinor[k2,MU,1]]]*ec3.ec4]
      +8*SUNSum[Col5,3]*SUNT[Glu3,Col5,Col2]*SUNT[Glu4,Col1,Col5]*mul[Alfas*MU*Pi]*
      abb[fme[WeylChain[DottedSpinor[k1,MU,-1],6,Spinor[k2,MU,1]]]*ec3.ec4
        -1/2*fme[WeylChain[DottedSpinor[k1,MU,-1],6,ec3,ec4,Spinor[k2,MU,1]]]
        +fme[WeylChain[DottedSpinor[k1,MU,-1],7,Spinor[k2,MU,1]]]*ec3.ec4
        -1/2*fme[WeylChain[DottedSpinor[k1,MU,-1],7,ec3,ec4,Spinor[k2,MU,1]]]] )
\end{verbatim}
\end{scriptsize}
while the abbreviated expression returning from Mathematica is just
\begin{scriptsize}
\begin{verbatim}
    -4*Den(U,MU2)*SUNSum(Col5,3)*SUNT(Glu3,Col5,Col2)*SUNT(Glu4,Col1,Col5)*
      AbbSum5*Alfas*Pi
\end{verbatim}
\end{scriptsize}


\subsection{Translation to Fortran code}

Numerical evaluation of the FormCalc results is typically done in 
Fortran, firstly for speed, and secondly for ease of inclusion into 
other programs.  The choice of Fortran by no means precludes usage in 
C/C++ as it is straightforward to invoke and link the Fortran code.
There is also a way to turn FormCalc-generated code into a 
Mathematica function \cite{mma}.

Code generation for the squared amplitude is a highly automated 
procedure.  FormCalc also offers low-level Fortran output functions with 
which it is very easy to turn an arbitrary Mathematica expression into 
Fortran code (see Ref.~\cite{excursions} for some examples of 
`non-standard' code generation).

\subsubsection{Code Generation for the Squared Amplitude}

FormCalc has two fairly advanced functions for generating 
Fortran code, \Code{WriteSquaredME} and \Code{WriteRenConst}.  The 
philosophy is that the user should not have to modify the generated 
code.  This means that the code has to be encapsulated (\ie no loose 
ends the user has to bother with), and that all necessary subsidiary 
files (include files, makefile) have to be produced, too.

First, a directory must be created for the code, and the driver programs 
copied into this directory with \Code{SetupCodeDir}.
\begin{alltt}
   dir = SetupCodeDir["\Var{fortrandir}"]
   WriteSquaredME[\Mtree,\,\Mloop,\,abbr,\,dir]
   WriteRenConst[\Mloop,\,dir]
\end{alltt}
\Code{WriteSquaredME} writes out a Fortran subroutine \Code{SquaredME} 
to numerically evaluate $|\Mtree|^2$ and $2\Re\Mloop\,\Mtree^*$, where 
\Mtree\ and \Mloop\ are outputs of \Code{CalcFeynAmp}.  
\Code{WriteRenConst} searches \Mloop\ for renormalization constants and 
writes out a subroutine \Code{CalcRenConst} for their computation.

The Fortran code is organized in a main code directory, which contains 
the main program and all its prerequisite files (\eg to set up 
kinematics and model parameters), and subsidiary `folders' 
(subdirectories to the main code directory).  The default setup looks 
like this:
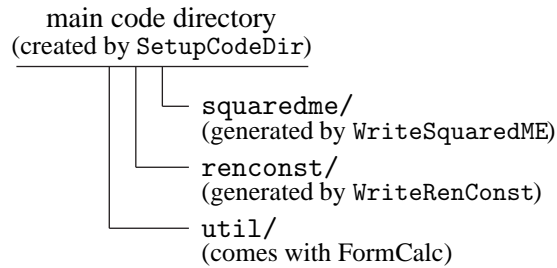
\begin{center}
\begin{picture}(200,100)
\Text(55,90)[b]{main code directory}
\Text(55,80)[b]{\small (created by \Code{SetupCodeDir})}
\Line(0,77)(110,77)
\Line(55,77)(55,62)
\Line(55,62)(65,62)
\Text(70,62)[l]{\Code{squaredme/}}
\Text(70,52)[l]{\small (generated by \Code{WriteSquaredME})}
\Line(45,77)(45,39)
\Line(45,39)(65,39)
\Text(70,39)[l]{\Code{renconst/}\hphantom{p}}
\Text(70,29)[l]{\small (generated by \Code{WriteRenConst})}
\Line(35,77)(35,16)
\Line(35,16)(65,16)
\Text(70,16)[l]{\Code{util/}\hphantom{p}}  
\Text(70,6)[l]{\small (comes with FormCalc)\hphantom{p}} 
\end{picture}
\end{center}
Each folder is equipped with its own makefile which makes a library of
the same name, \eg the makefile in \Code{util/} makes the library
\Code{util.a}.  These sub-makefiles are orchestrated by the master
makefile.

Occasionally it is useful to have more than one instance of 
\Code{squaredme} (or \Code{renconst}), \eg when computing an hadronic 
cross-section to which several partonic processes contribute. Both 
\Code{WriteSquaredME} and \Code{WriteRenConst} have the \Code{Folder} 
option, with which a unique folder name can be chosen, and the 
\Code{SymbolPrefix} option, with which the symbols visible to the linker 
can be prefixed with a unique identifier.


\subsubsection{Low-level code-generation functions}

FormCalc's code-generation functions are also publicly available.  They 
can be used to write out an arbitrary Mathematica expression as 
optimized Fortran code.  The basic procedure is simple:
\begin{enumerate}
\itemsep=0pt
\item
\Code{\Var{handle} = OpenFortran["\Var{file.F}"]} \\
opens \Var{file.F} as a Fortran file for writing,   

\item
\Code{WriteExpr[\Var{handle},\ \{\Var{var} \to \Var{expr},\,\dots\}]} \\
writes out Fortran code to calculate \Var{expr} and store the
result in \Var{var},

\item
\Code{Close[\Var{handle}]} \\
closes the file again.
\end{enumerate}
The code generation is fairly sophisticated and goes well beyond
merely applying Mathematica's \Code{FortranForm}.  The generated code 
is optimized, \eg common subexpressions are pulled out and computed in 
temporary variables.  Expressions too large for Fortran are split into 
parts, as in
\begin{verbatim}
     var = part1
     var = var + part2
     ...
\end{verbatim}
If the expression is too large even to be reasonably evaluated in one 
file, \eg if the compile time becomes too long, the \Code{FileSplit} 
function can distribute it on several files and optionally write out a 
master subroutine which invokes the individual parts.

To further automate the code generation, such that the resulting code 
needs few or no changes by hand, many ancillary functions are available, 
\eg \Code{VarDecl} writes out variable declarations for a given list of 
variables.


\section{LoopTools}

LoopTools supplies the actual numerical implementations of the one-loop 
integrals needed for programs made from the FormCalc output.  It is 
based on the FF package \cite{FF} and provides in addition to the 
scalar integrals of FF also the tensor coefficients in the conventions 
of \cite{De93}.  LoopTools offers three interfaces: Fortran, C++, and 
Mathematica.

Using the LoopTools functions in Fortran and C++ is very similar.  In 
Fortran it is necessary to include the file \Code{looptools.h} in every 
function or subroutine (for the common blocks).  In C++, 
\Code{clooptools.h} must be included once.  Before using any LoopTools 
function, \Code{ffini} must be called and at the end of the calculation 
\Code{ffexi} may be called to obtain a summary of errors.

Recent additions to LoopTools include:
\begin{itemize}
\item The remaining complex-mass case of a $D_0$ with four complex 
masses has been added \cite{D0c}.

\item
Dimensionally regulated IR and collinear divergences, so far implemented 
publicly only in QCDLoop \cite{QCDLoop} have been added.  Currently 
only the scalar integrals are available (as in QCDLoop).

Technically, the parameter $\lambda^2$ (\Code{LTLAMBDA}), used hitherto 
to set the IR regulator (`photon') mass and thus implicitly assumed 
positive, now includes the cases $\lambda^2 = -2$, $\lambda^2 = -1$, and 
$\lambda^2 = 0$, in which cases the $1/\varepsilon^2$, $1/\varepsilon$, 
and finite piece are returned.

\item
The dispatcher for IR and collinear divergences has been replaced by a 
more efficient code.  It constructs a bit pattern: 1 for zero argument,
0 otherwise, and can then jump to the correct case with a single table 
lookup.
\end{itemize}

\section{Requirements and Availability}

FeynArts, FormCalc, and LoopTools are available from
\begin{verbatim}
   http://feynarts.de
   http://feynarts.de/formcalc
   http://feynarts.de/looptools
\end{verbatim}
This website carries a script \Code{FeynInstall} for easy installation.  
Each package contains a comprehensive manual.  All three packages are 
open source and licensed under the LGPL.

\end{document}